\documentstyle[12pt]{article}
\def\be{\begin{equation}}
\def\ee{\end{equation}}
\def\bear{\be\begin{array}}
\def\eear{\end{array}\ee}
\def\bea{\begin{eqnarray}}
\def\eea{\end{eqnarray}}

\def\baselinestretch{1}
\begin{document}
\catcode`@=11
\newtoks\@stequation
\def\subequations{\refstepcounter{equation}%
\edef\@savedequation{\the\c@equation}%
  \@stequation=\expandafter{\theequation}
  \edef\@savedtheequation{\the\@stequation}
  \edef\oldtheequation{\theequation}%
  \setcounter{equation}{0}%
  \def\theequation{\oldtheequation\alph{equation}}}
\def\endsubequations{\setcounter{equation}{\@savedequation}%
  \@stequation=\expandafter{\@savedtheequation}%
  \edef\theequation{\the\@stequation}\global\@ignoretrue

\noindent}
\catcode`@=12
\begin{titlepage}

 \title{STRING PHENOMENOLOGY \footnote{Based on a talk given at
"Pacific Particle Physics Phenomenology", Seoul (S. Korea), 
31 October-2 November 1997.} 
}
 \author{{ C. MU\~NOZ}
\\
\hspace{3cm}\\
{\small Departamento de F\'{\i}sica 
Te\'orica C--XI}\\ 
{\small Universidad Aut\'onoma de Madrid,
Cantoblanco, 28049 Madrid, Spain} \\
{\small e--mail: cmunoz@delta.ft.uam.es}}
\date{}
\maketitle
\def\baselinestretch{1.15}
\begin{abstract}
\noindent
Following Dyson's analogy
between the quantum field theory and the 19th-century chemistry--both
explain {\it how} but not {\it why}--one could also establish an
analogy between atomic physics and string theory. Atomic physics
was needed to answer the question {\it why} in chemistry,
string theory is trying to answer the question {\it why}
in the standard model. This attempt is very briefly reviewed in this 
talk. After discussing the phenomenological aspects of string theory,
the study of soft terms is emphasized as a possible connection between
theory and experiment. Special attention is paid to the issues of
flavor changing neutral currents, loop effects, charge and color
breaking minima, and M--theory versus
weakly coupled string theory.

\end{abstract}

\thispagestyle{empty}

\leftline{}
\leftline{}
\leftline{FTUAM 98/2}
\leftline{February 1998}


\end{titlepage}
\newpage
\setcounter{page}{1}

\section{Introduction}
Dyson, in an entertaining article written in 1953 \cite{Dyson},
drew the following analogy between the quantum field theory and
the
19th--century chemistry: 
"The latter described the properties of the chemical elements and their
interactions. How the elements behave; it did not try to explain why
a particular set of elements, each with its particular properties, exists.
To answer the question {\it why}, completely new sciences were needed:
atomic and nuclear physics. (...)
The quantum field theory treats elementary particles just as 19th--century
chemists treated the elements. The theory is in its nature descriptive and
not explanatory. It starts from the existence of a specified list of
elementary particles, with specified masses, spins, charges and specified
interactions with one another. All these data are put into the theory
at the beginning. The purpose of the theory is simply to deduce from this
information what will happen if particle $A$ is fired at particle $B$ with
a given velocity. We are not yet sure whether the theory will be able to 
fulfill even this modest purpose completely. Many technical difficulties
have still to be overcome. One of the difficulties is that we
do not yet have the complete list of elementary particles. Nevertheless
the successes of the theory in describing experimental results have been
striking. It seems likely that the theory in something like its present
form will describe accurately a very wide range of possible experiments.
This is the most that we would wish to claim for it".

Now, in 1998, we do have the complete list of elementary particles (at 
least at energies below the electroweak scale, and given some caveats 
related with the Higgs sector)--the proton, pi meson, etc.
which appeared in the chart of fundamental particles in Dyson's presentation
should be regarded as an amusing historical anecdote--and
we do know that the theory fulfilling the modest purpose mentioned by 
Dyson is the {\it standard model}. 
What the 19th--century chemistry did with the chemical elements,
the 20th--century standard model does with the elementary particles.
It describes how the elementary particles behave but does not try to 
explain why a particular set of elementary particles, each with
its particular properties, exists. To answer the question {\it why} it 
seems
that new sciences are not needed, as atomic and nuclear physics in the
case of chemistry, but 
just new theories. This is the precisely the purpose of the 
{\it string theory}.

\section{String Phenomenology}

In string theory the elementary particles are not point--like objects but
extended, string--like objects. It is surprising that this 
apparently 
small change may allow us to answer fundamental questions that in the 
context
of the quantum field theory of point--like particles cannot even be posed:
why do we live in four dimensions?, why is the gauge group
$SU(3)\times SU(2)\times U(1)$?, why are there three families of particles?,
why is the pattern of quark and lepton masses so strange?,
why is the fine structure constant given by $\alpha=1/137$?, etc.
For example, with respect to the last two questions, let us remark that in 
string theory there are no free parameters, all
coupling constants are in fact no constants but {\it fields}. In particular
the measured experimental values of the gauge coupling constants
and Yukawa couplings correspond to the vacuum expectation values (VEVs)
of those fields.  
The gauge singlet field $S$, called dilaton, determines
the gauge couplings whereas the gauge singlet fields $T_i$, called
moduli, determine the Yukawa couplings. These fields are generically
present in four--dimensional models: the dilaton arises from the 
gravitational
sector of the theory and the moduli parametrize the size and shape of the
compactified variety (let us recall that 
string theory can only be quantized 
in ten dimensions). Although the most phenomenologically promising string 
theory, the heterotic string, has a gauge group $E_8\times E_8$ in ten
dimensions, there exist different methods 
to 
build standard--like models (i.e. gauge group 
$SU(3)\times SU(2)\times U(1)$ with three families) in four 
dimensions \cite{Quevedo}:
orbifolds, Calabi--Yau spaces,
fermionic strings, covariant lattices, Gepner models, etc.
In spite of these achievements it is fair to say that
the standard model has not been obtained yet. In this sense the problem
of string theory is that it is too ambitious: the correct model
must reproduce not only the gauge group and families of the standard model
but also the correct values of the gauge coupling constants,
the correct mass hierarchy for quark and leptons, a realistic 
Kobayashi--Maskawa matrix, etc., i.e. the nineteen parameters fixed
by the experiment in the standard model. In order to 
reach such a goal thousands of models 
(vacua) can be built due to the fact that there are many 
consistent
ways to compactify the extra dimensions.
Many models have 
a number of families different from three, 
no appropriate gauge groups, no appropriate matter, etc.,
but others
have the gauge group of the standard model or grand unified gauge groups,
and three families of particles. 
Then, using the experimental results available (e.g. fermion
masses or mixing angles), to discard most of the models in order to 
obtain the standard model is possible in principle \cite{nosotros}. 
Unfortunately, the
model space is so hughe that this type of analysis is very cumbersome in
practice.
Another possibility is to select the correct model (vacuum) using some
(unknown for the moment) 
{\it dinamical} mechanism. In this sense duality symmetries may help to solve
the problem. The five consistent string theories, type I with gauge 
symmetry $SO(32)$, heterotics with gauge symmetries $E_8\times E_8$ or 
$SO(32)$, type IIA and type IIB, are in fact connected by 
dualities \cite{quien} 
and may
be the thousands of vacua of these theories are also connected. Thus
due to these connections some mechanism could select one point in the
hughe model space. Besides, it has recently been proposed that the five 
string theories can be derived from a single theory, called 
M--theory \cite{quien}.
This might be the underlying fundamental theory of elementary particles.
For example, the strong
coupling limit of the ten--dimensional $E_8\times E_8$ heterotic string 
may be obtained compactifying the eleven--dimensional M--theory on an 
orbifold. The resulting theory, unlike the weakly coupled heterotic string,
provides an attractive framework for the unification of 
couplings \cite{Dienes}.
Although the quantum version of M--theory is yet to be built, 
some properties are already known. It is related with {\it membranes} and 
its low--energy limit is {\it eleven}--dimensional supergravity (SUGRA).


%
\section{"Soft" Phenomenology 
}

We saw in the previous section that to obtain a connection between theory
and experiment (i.e. the standard model) in strings is possible in principle
but difficult in practice. Here we will discuss a possible way--out to this
problem.
If Nature is supersymmetric (SUSY) at the weak scale, 
as many particle physicists
believe, eventually the spectrum of SUSY 
particles will be measured providing us with a possible connection 
with the superhigh--energy world of 
string theory.
The point is that the SUSY spectrum is determined by
the soft SUSY--breaking parameters which 
can be computed in principle in the context of (super)string models.
To compare then the superstring theory predictions about soft terms 
with
the experimentally observed SUSY spectrum would allow us 
to test the goodness of this theory.
This is the path, followed in the last years by several groups, 
that I will try to review very briefly in this sect. 
I will
concentrate in subsect. 3.1 on weakly coupled heterotic string models
leaving the discussion of
strongly coupled strings (from M--theory) 
for subsect. 3.2.


\subsection{From Superstrings}


As is well known the soft parameters, scalar masses $m_{\alpha}$,
gaugino masses $M_a$, trilinear $A_{\alpha\beta\gamma}$ and bilinear 
$B_{\alpha\beta}$ 
coefficients, can be computed in generic hidden
sector SUGRA models \cite{BIM3}.
They depend on the three functions,
which determine the full N=1
SUGRA Lagrangian:
the K\"ahler potential $K$, the superpotential $W$,
and the gauge kinetic function $f$. 
However one can think of many possible
SUGRA models (with different $K$, $W$ and $f$) leading to
{\it different} results for the soft terms. This arbitrariness can be 
ameliorated in SUGRA
models deriving from superstring theory, where $K$, $f$, and the hidden
sector are more constrained. They have
a natural hidden sector built--in: the dilaton
field $S$ and the moduli fields $T_i$.
The associated effective SUGRA K\"ahler potentials, at the superstring
tree level, 
are of the type: 
\begin{eqnarray}
K &=& -ln(S+\overline S)
+{\hat K}(T_i,{\overline T}_i)
+ {\tilde K}_{{\alpha}{\overline{\beta}}}
C^{\alpha} { \bar C^{\overline{\beta}} }\ 
+\frac{1}{4}
{\tilde K}_{{\alpha}{\overline{\beta}}\gamma{\overline {\delta}}}
C^{\alpha} \bar C^{\overline{\beta}}
C^{\gamma} \bar C^{\overline{\delta}}
\nonumber\\
&& 
+
\left[ \frac{1}{2} Z_{{\alpha }{ \beta }}
C^{\alpha}C^{\beta } 
+\frac{1}{2}
{Z}_{{\alpha}\beta{\overline{\gamma}}}
C^{\alpha} C^{\beta} \bar C^{\overline{\gamma}}  
\right.
\nonumber\\
&&
\left. + \frac{1}{6} Z_{{\alpha }{ \beta }\gamma\overline{\delta}}
{C}^{\alpha}
C^{\beta } C^{\gamma} \bar C^{\overline{\delta}} 
+\ h.c.   \right]
+... 
\label{kahler}
\end{eqnarray}
where the coefficient functions depend upon $T_i$, ${\overline T}_i$
and the ellipsis denote the terms which are irrelevant for the present
calculation.
The first piece is the usual term corresponding to the complex dilaton
$S$ that is present for any compactification. The second piece is the
K\"ahler potential of the moduli fields, which in general depends on the
compactification scheme and can be a complicated function. For the moment
we leave it generic, but in the following we will analyze some specific 
classes of superstring models where it has been computed. The same
comment applies to the coefficient functions ${\tilde K}_{\alpha \overline 
{\beta}} (T_i,{\overline T}_i)$,
${Z}_{\alpha \overline {\beta}}
(T_i,{\overline T}_i)$, etc. Finally, for any four--dimensional
superstring the tree--level gauge kinetic function is independent of the
moduli sector and is simply given by 
\begin{equation}
f_{a}= k_a S\ , 
\label{kinetic}
\end{equation}
where $k_a$ is the Kac--Moody level of the gauge factor.
On the other 
hand, the superpotential $W$ is more involved since 
it is related with the mechanism of SUSY breaking.
For example, non--perturbative gaugino condensation is an interesting
mechanism giving rise to an effective superpotential $W(S,T_i)$
which breaks SUSY at the same time that $S$ and $T_i$ acquire
reasonable VEVs. The structure of soft parameters can be analyzed and
in particular, at low--energies, squarks are almost degenerate with
sleptons and much heavier than gluinos \cite{Beatriz}. However, one 
should be cautious
about this result due to the fact that the cosmological constant generated
by this mechanism is non--vanishing.

On the other hand,
without specifying the SUSY--breaking mechanism, 
just
assuming that the auxiliary
fields of $S$ and $T_i$, $F^S$ and $F^i$ respectively, are the seed of 
supersymmetry breaking, still interesting
predictions for this simple class of models can be obtained \cite{BIM3}.
This is the approach that we will discuss here.

Let us focus first on the very interesting limit where the dilaton $S$
is the source of all the SUSY breaking. At superstring tree level
the dilaton couples
in a universal manner to all particles and therefore, this limit is 
compactification {\it independent}. In particular, since  
${\tilde K_{{\alpha}{\overline{\beta}}}}
(T_i,{\overline T}_i)$ in 
(\ref{kahler})
is independent on $S$,
the soft scalar masses turn out to be 
universal
$m_{\alpha}=m_{3/2}$.
Because of the simplicity of this scenario \footnote{Modifications to this 
scenario due to the effect of possible superstring
non--perturbative corrections to the K\"ahler potential have also
been analyzed \cite{alberto}.},
the predictions about 
the low--energy spectrum 
are quite precise 
\cite{BIM3}. 
For example, for the first two generations,
squarks are almost degenerate with
gluinos and much heavier than sleptons.

In general the moduli fields, $T_i$, 
may also contribute to SUSY breaking,
and in that case their effects on soft parameters must also be 
included.
Since different compactification schemes give rise to different
expressions for the moduli--dependent part of $K$
(\ref{kahler}), 
the computation of the soft parameters will be 
model {\it dependent}. 
To illustrate the main features of mixed dilaton/moduli--dominated 
SUSY breaking, let us 
concentrate first on the simple situation of diagonal moduli and matter 
metrics. This is the case of most
$(0,2)$ symmetric Abelian orbifolds which
have
${\tilde K}_{\alpha{\overline{\beta}}}=\delta_{\alpha\beta}
\Pi_i(T_i+{\overline T}_i)^{n_{\alpha }^i}$
where $n_{\alpha }^i$ are the 
modular weights of the matter fields $C^{\alpha }$. 
Then e.g. the scalar masses 
are given by:
\begin{eqnarray}
{m}^{2}_{\alpha} = 
m_{3/2}^2+\sum_i \frac{n_{\alpha}^i}{(T_i+ {\overline{T}_i})^2}
|F^i|^2 
\label{supermasss}
\end{eqnarray}
With this information one can analyze the structure of soft parameters
and in particular the low--energy spectrum \cite{BIM3}.
Although 
continuous Wilson--line moduli may also contribute to
SUSY breaking, 
the results turn out to be similar
\cite{kim}. 

Universality is a desirable property for phenomenological reasons, 
particularly to avoid flavor changing neutral currents (FCNC). 
Notice however that the scalar masses (\ref{supermasss}) 
show in general a {\it lack of universality} due to the
modular weight dependence \cite{lust}. 
So, even with diagonal matter metrics, FCNC effects may appear. 
%
%
Of course, the same situation will appear for the few orbifolds,
$Z_3$, $Z_4$ and
$Z'_6$, where off--diagonal metrics are present in the untwisted sector
giving rise to a generic matrix structure for the scalar masses
$m_{\alpha\overline{\beta}}$ \cite{BIMS}.
This potential problem seems to be endemic in superstring models \cite{yo}.
E.g.  
in Calabi--Yau compactifications off--diagonal metrics is the generic
situation and therefore 
the mass eigenvalues are 
typically non--degenerate \cite{kim2}. 

In fact, this situation may even be aggravated once one includes loop 
effects. On the one hand, superstring loop effects may spoil the
universality in the dilaton--dominated SUSY--breaking limit due
to an $S$--dependent contribution in 
${\tilde K}_{\alpha \overline {\beta}}$ (\ref{kahler}) \cite{louis}.
On the other hand,
the soft parameters 
are usually computed
at the tree level of SUGRA interactions. However, as has been 
shown explicitly \cite{choi}, 
there can be a significant modification in this procedure
due to quadratically divergent SUGRA one--loop effects.
In particular,
due to the contributions
of the pieces
depending upon 
${Z_{{\alpha}\beta{\overline{\gamma}}}}$ and 
${\tilde K_{{\alpha}{\overline{\beta}}\gamma{\overline{\delta}}}}$
in $K$ (\ref{kahler}),
the scalar masses will have a generic matrix structure with 
non--degenerate eigenvalues \footnote{
It is worth noticing here that in the case of the $\mu$ term
these quantum corrections 
do not only modify the already known contributions 
from $\tilde{\mu}_{\alpha\beta}$ and $Z_{\alpha\beta}$
in the tree--level matching condition \cite{BIM3}
but also provide {\it new} sources of the $\mu$ term, naturally of order
the weak scale \cite{choi}. These sources 
depend upon the coefficients
$Z_{\alpha\beta\overline{\gamma}}$ and 
$Z_{\alpha\beta\gamma\overline{\delta}}$ in $K$
and also possible coefficients $\tilde{f}_{a\alpha\beta}$
in $f_{a}$.
}.
Let us remark that 
results from loop effects by SUGRA interactions 
are valid
in general, and 
therefore 
can be applied to any superstring theory, and in particular to 
M--theory.

Nevertheless,
we recall that the low--energy running of the scalar masses has to be
taken into account. In particular, in the squark case, for gluino masses 
heavier than (or of the same order as) the scalar masses at the boundary
scale, there are large flavour--independent gluino loop contributions 
which are the dominant source of scalar masses.
This situation is very common e.g. in orbifold models.
The above effect can therefore help in fulfilling the FCNC 
constraints \cite{BIM}.


Let us finally recall that
another type of constraints arise from
demanding the no existence of
low--energy
charge and color breaking minima deeper than the standard vacuum 
\cite{mua}. These restrictions are in general very strong and in
the particular
case of the
dilaton--dominated scenario
the whole parameter space ($m_{3/2}$, $B$, $\mu$)
turns out to be excluded on these grounds \cite{mua2}.

\subsection{
From M--Theory}

Given the theoretical and phenomenological virtues of M--theory mentioned
in sect. 2, it is crucial to examine how the pattern of soft terms changes
when one moves from the weakly coupled heterotic string limit to the
M--theory limit. 
For example, 
higher order corrections to the K\"ahler potential in the M--theory limit
imply a large $S$--dependent contribution in 
${\tilde K}_{\alpha \overline {\beta}}$ (\ref{kahler}) \cite{ovrut}.
Under the same assumption than in the previous subsect., i.e. that SUSY
is spontaneously broken by the auxiliary components of the bulk $S$ and $T_i$
superfields, it has been shown explicitly \cite{mtheory}
that there can be a sizable difference between the heterotic string limit
and the M--theory limit even in the overall pattern of soft terms.
With respect to the issue of FCNC,
due to the above mentioned $S$--dependent contribution, there
can be a large violation of the scalar mass universality even in the
dilaton--dominated scenario. Notice that a similar effect commented in
the previous subsect. is expected to be much smaller since it is due to 
string loops.

\section{Final comments}
Let us hope that the comment about chemists that Dyson wrote in the 
article
mentioned in the introduction--"Looking backward, 
it is now clear that 19th--century chemists were
right to concentrate on the {\it how} and to ignore the {\it why}.
They did not have the tools to begin to discuss intelligently
the reasons for the individualities of the {\it elements}. They had to
spend a hundred years building up a good quantitative descriptive
theory before they could go further. And the result of their labors, 
the
classical {\it science of chemistry}, was not destroyed or superseded by the 
later insight that {\it atomic physics} gave."--will 
also be written by somebody in the next century
about 20th--century physicists substituting elements by {\it 
elementary 
particles}, science of chemistry by {\it standard model} and 
atomic 
physics by {\it string theory}.

\section*{Acknowledgments}
Research supported in part by: the CICYT, under contract AEN97-1678-E;
the European Union, under contract CHRX-CT93-0132.


\end{document}